# Urban vs. rural divide in HTTPS implementation for hospital websites in Illinois.

Robert Robinson
Department of Internal Medicine
SIU Medicine
Springfield, IL, USA
rrobinson@siumed.edu

## Abstract

The Hypertext Transfer Protocol Secure (HTTPS) communications protocol is used to secure traffic between a web browser and server.  This technology can significantly reduce the risk of interception and manipulation of web information for nefarious purposes such as identity theft.  Deployment of HTTPS has reached about 50% of all webs sites.  Little is known about HTTPS implantation for hospital websites.  To investigate the prevalence of HTTPS implementation, we analyzed the websites of the 210 public hospitals in the state of Illinois, USA.  HTTPS was implemented to industry standards for 54% of all hospital websites in Illinois.  Geographical analysis showed an urban vs. rural digital divide with 60% of urban hospitals and 40% of rural hospitals implementing HTTPS.

## Introduction

Properly configured Hypertext Transfer Protocol Secure (HTTPS) communications protocols secure websites by providing confidentiality, authentication, and integrity to the website.  This technology can significantly reduce the risk of digital eavesdropping, content modification, and redirection when visiting a website.  HTTPS is highly desirable for healthcare related websites because of the personal nature of healthcare and concerns about cybercrimes such as identity theft (Finley 2017, General Services Administration and the CIO Council, Electronic Frontier Foundation).

National and international concerns regarding pervasive surveillance, cybercrime, and cyberwarfare have prompted increasing adoption of HTTPS technology for many websites.  By January of 2017, half of all web traffic for common web browsers went to HTTPS secured websites (Gebhart 2017, Finley 2017, Electronic Frontier Foundation 2017).  Two-thirds of high traffic federal government websites implemented HTTPS by November 2017 (Castro, Nurko and McQuinn 2017) with a goal of making all sites HTTPS-only in the near future (General Services Administration and the CIO Council).  HTTPS implementation was completed for 51% of all federal government websites by February 2018 (General Services Administration, 2018).

To investigate the implementation of this important information security measure at healthcare institutions, this study will determine the prevalence of HTTPS implementation on the websites of public hospitals in the state of Illinois.

## Methods

This study used the HTTPS evaluation methodology described by Castro, Nurko and McQuinn for Benchmarking U.S. Government Websites (Castro, Nurko and McQuinn 2017).

The Illinois Department of Public Health (IDPH) hospital directory was downloaded from the IDPH website [https://data.illinois.gov/dataset/410idph_hospital_directory].  This directory included all 210 licensed public hospitals in the state of Illinois as of September 27, 2017.

Each listing includes the hospital legal name, city and county the hospital is located in and type of hospital.   Hospitals in counties designated as rural by the US Office of Management and Budget were identified as rural hospitals.

IDPH defined hospital types are:  general hospital, critical access hospital (no more than 25 inpatient beds), long-term acute care hospital (a hospital focused on patients who have an average hospital stay of more than 25 days), pediatric hospital (exclusively serves children), psychiatric hospital (a hospital focused on psychiatric care), and rehabilitation hospital (a hospital focused on rehabilitation after stabilization of acute medical issues).

For each hospital on the list, the Google search engine (https://www.google.com) was used to identify the hospital website by searching for the hospital name, city, and state.  The hospital web site uniform resource locator (URL) was entered into the SSL/HTTPS evaluation tool (https://www.ssllabs.com/ssltest/) developed by Qualys for analysis.

The evaluation tool performs a validation on the HTTPS configuration for the entered URL, returning a 'grade' of A through F based on the implementation of standards of the certificate, protocol support, key exchange, cipher strength, and specific vulnerabilities.  A grade of "A" with an overall score of 90 or better was required to pass the HTTPS implementation test.

These tests were performed on December 11, 2017.

The research protocol was reviewed by the Springfield Committee for Research Involving Human Subjects, and it was determined that this project did not fall under the purview of the IRB as research involving human subjects according to 45 CFR 46.101 and 45 CFR 46.102.

## Statistical analysis

Qualitative variables were compared using Pearson chi2 or Fisher's exact test and reported as frequency (%).   Statistical analysis was performed using SPSS version 22 (SPSS Inc., Chicago, IL, USA).

## Results

Websites could be identified for all 210 hospitals in the study sample.

HTTPS implementation could be found for 76% of Illinois hospital websites with 54% passing the HTTPS evaluation (Table 1). HTTPS implementation was much more common for urban than rural hospital websites (84% vs. 55%, p < 0.001). Passing the HTTPS evaluation was more common in urban than rural hospital websites (60% vs. 40%, p = 0.009).

Most hospitals in rural areas are characterized as critical access hospitals (62% vs. 10%, p < 0.001) and general hospitals are more common in urban areas (73% vs. 38%, p < 0.001). Long term acute care, pediatric, psychiatric, and rehabilitation hospitals were not found in rural areas.

Table 1. Comparison of HTTPS implementation in urban and rural hospitals in Illinois

|  | Urban Hospitals N = 152 | Rural Hospitals N = 58 |  |
|---|---|---|---|
| **HTTPS Implemented** | 127 (84%) | 32 (55%) | **p < 0.001** |
|  |  |  |  |
| **HTTPS Implementation** |  |  |  |
| Pass (Grade A) | 91 (60%) | 23 (40%) | **p = 0.009** |
| Fail (Grades B, C, F) | 36 (24%) | 9 (16%) | p = 0.197 |
|  |  |  |  |
| **Hospital Type** |  |  |  |
| General | 111 (73%) | 22 (38%) | **p < 0.001** |
| Critical Access | 15 (10%) | 36 (62%) | **p < 0.001** |
| Long Term Acute Care | 9 (6%) | 0 (0%) | p = 0.058 |
| Pediatric | 3 (2%) | 0 (0%) | p = 0.281 |
| Psychiatric | 10 (7%) | 0 (0%) | **p = 0.045** |
| Rehabilitation | 4 (3%) | 0 (0%) | p = 0.212 |

## Discussion

HTTPS implementation is common (76% overall) for hospital websites in Illinois, but only 54% pass the HTTPS implementation test. This result is less than the 67% rate of passing the HTTPS implementation test for federal government websites (Castro, Nurko and McQuinn 2017) but comparable to the 50% rate of HTTPS implementation for all websites (Gebhart 2017, Finley 2017).

Given the potential value of improving the confidentiality, authentication, and integrity of the information on a hospital website to the general public, these results show a much lower than expected rate of HTTPS implementation.

The reasons for this lower rate of passing the HTTPS implementation evaluation for Illinois hospital websites is unknown. However, failing to appropriately implement HTTPS may put hospital website visitors at risk of redirection to a fraudulent website, data interception, and data manipulation using techniques successfully used against visitors to other websites.

HTTPS implementation for hospital websites in Illinois shows a distinct urban vs. rural digital divide (60% vs. 40% HTTPS implementation, p = 0.009) that mirrors other assessments of the digital divide in the United States (Carlson and Goss 2016; Perrin 2017; Hale, Cotton, Drentea and Goldner 2010).

Rural America also has a healthcare divide.  A systematic review shows that people living in rural communities have reduced access to healthcare services that is compounded by factors such as poor public transport, scarcity of healthcare resources (physicians, hospitals, clinics), and higher costs (Douthit, Kiv, Dwolatzky, and Biswas.  2015). Many of these factors are further complicated with reduced access to broadband internet services which could improve the availability and efficiency of existing health services in rural areas.

## Conclusions

HTTPS implementation for hospital websites in Illinois mirrors the rate for all websites (50%).  The geography of HTTPS implementation for hospital websites in Illinois shows an urban vs. rural divide seen in other assessments on the availability of internet and healthcare services.  Further research is needed to better understand the significance of this data and the potential implications for the healthcare system.